\magnification=\magstep1
\hfuzz= 6pt
\baselineskip= 16pt

$ $

\vfill

\centerline{\bf Almost Certain Escape from Black Holes}

\vskip 1cm

\centerline{Seth Lloyd$^*$}

\centerline{ MIT Mechanical Engineering}

\vskip .5 cm

\noindent{\it Abstract:}  
Recent models of the black-hole final state suggest that quantum information can
escape from a black hole by a process akin to teleportation.  
These models require a specific final state and restrictions
on the interaction between the collapsing matter and the incoming
Hawking radiation for quantum information to escape.  This paper
investigates escape from black holes for arbitrary final states 
and for generic interactions between matter and Hawking radiation. 
Classical information, including the result of any computation performed by the
matter inside the hole, escapes from the hole with certainty.  
Quantum information escapes with fidelity $\approx (8/3\pi)^2$:
only half a bit of quantum information is lost on average, 
independent of the number of bits that escape from the hole.

\vskip 1in

It has been proposed that black holes could function as quantum
computers [1-2]; the computational capacity of black holes can
be calculated in terms of their mass and lifetime [1-3]. 
In order to function
as a useful computer, however, a black hole must permit information
to escape as the black hole evaporates.  Recently, Horowitz and Maldacena 
proposed a model of black hole evaporation that
imposes a final state boundary condition at the black-hole
singularity [4].  The result is a nonlinear time evolution for the quantum
states in and outside of the black hole, which permits quantum information
to escape from the black hole by a process akin to teleportation.  
Because it allows information to escape,
such a model naturally allows the black hole to function as a computer
whose output is written in the outgoing Hawking radiation produced
during evaporation, as envisioned in [1].  The Horowitz-Maldacena model
requires a specific final state which is perfectly entangled between
the matter that formed the black hole and the incoming Hawking radiation.
Whether or not quantum gravity supports such a final state remains to 
be seen.  In addition, even with the proper final state, interactions between 
the incoming Hawking radiation and the collapsing matter can spoil the 
unitary nature of the black-hole evaporation [5], destroying
some or all of the quantum information inside the hole [6-7].  

The purpose of this paper is to examine the robustness of the escape
of quantum information during black hole evaporation in final
state projection models. In particular, I show that for projection
onto {\it any} final state at the singularity
(independent of the exact details of quantum gravity) and for {\it almost
all} interactions between the matter and incoming Hawking radiation,
properly encoded classical information escapes from the hole with certainty.   
Of the quantum information that escapes from the hole, only one half
a qubit is lost on average, regardless of the number of bits of
quantum information in the hole to begin with.  More precisely,
the state of the matter that formed the hole
is preserved under black hole evaporation with a fidelity of
$f \approx (8/3\pi)^2 \approx .85$.  This is the fidelity of escape
of the entire state of the collapsing matter: individual quantum
bits escape with a fidelity that approaches one as the number of
bits in the hole becomes large.  Since the fidelity is above
the threshold for the use of quantum-error correcting codes,
properly encoded quantum information can escape from the hole with 
fidelity arbitrarily close to 1. 
  
The Horowitz-Maldacena model [4] is described concisely in [5].  
Black holes evaporate by absorbing negative-energy
`incoming' Hawking radiation and by emitting positive-energy `outgoing'
Hawking radiation.  
Let the dimension of the Hilbert space for the collapsing 
matter inside the black hole be $N$.  
In the ordinary semi-classical treatment of 
black-hole evaporation, the incoming and outgoing Hawking radiation 
is in the maximally entangled state
$$|\phi\rangle_{in\otimes out} = {1\over \sqrt N} 
\sum_{j=1}^N|j\rangle_{in} \otimes |j\rangle_{out}, \eqno(1) $$
where $\{ |j\rangle_{in} \}$ is an orthonormal basis for the Hilbert
space $H_{in}$ of the incoming
Hawking radiation and $\{ |j\rangle_{out} \}$ is an orthonormal basis 
for the Hilbert space $H_{out}$ of the outgoing Hawking radiation.

Let $|\phi\rangle_{matter\otimes in} \in H_{matter} \otimes H_{in} $ 
be the final state onto which the collapsing matter together with the 
incoming Hawking radiation is projected at the singularity.  
Horowitz and Maldacena postulated a form for this state of
$$|\phi\rangle_{matter\otimes in} = {1\over \sqrt N}
\sum_{k=1}^N (S |k\rangle_{matter}) \otimes |k\rangle_{in}, \eqno(2) $$
where $S$ is a unitary transformation acting on the matter states alone.  
The usual analysis of quantum
teleportation shows that for states of this form,
the transformation from the state of the collapsing matter
to the state of the outgoing Hawking radiation is 
$$ T= { }_{matter\otimes in}\langle \psi| \phi\rangle_{in\otimes out} 
= S/N. \eqno(3)$$  
The factor $1/N$ reflects the fact that if
this were conventional teleportation, then this particular final state
would occur only with probability $1/N^2$.  In final-state projection
however, only one final state can occur: accordingly,
the final transformation from collapsing matter to outgoing Hawking
radiation is renormalized, and the net result is the unitary transformation 
$S$. In the H-M model, final-state projection leads to a unitary transformation
between collapsing matter and the outgoing Hawking radiation.

Comparing final-state projection to conventional teleportation, we see
that the main difference is that final-state projection mandates a single
outcome, while teleportation allows $N^2$ outcomes.  In teleportation,
$ \log_2 N^2$ bits must be sent from the input of the teleporter to its
output in order to reconstruct the input state.  In final-state
projection $\log_2 1 = 0$ bits must be sent from inside the black
hole to outside the black hole to reconstruct the input state.    

Final state projection is an intrinsically nonlinear process and
shares the virtues and vices of other proposals for nonlinear
quantum mechanical processes.  Escape
of quantum information from black holes via final state projection is 
similar to the use of nonlinear quantum mechanics to 
provide superluminal communication as described (and rejected) in [8-10], 
to violate the second law
of thermodynamics [11], or to solve NP-complete problems [12].
Such nonlinear quantum effects have been investigated experimentally
under non-Planckian conditions and ruled out to a high degree of
accuracy [13-16] ([6-7] proposes similar tests of such nonlinear
quantum effects in a `normal' environment).  In fact, the nonlinearity
that arises from projection onto a state is a particularly powerful
type, capable of allowing superluminal communication and time travel
if it occurs under normal conditions.  Indeed, this nonlinearity
must allow information to propagate over spacelike intervals
if the information is to escape from a black hole.   Because it
occurs at a singularity beyond an event horizon, however, final state projection
does not obviously allow causal paradoxes such as time travel.  

Despite its somewhat dubious provenance, nonlinear quantum mechanics including
final state projection might hold sway in extreme Planckian regimes
such as the black-hole singularity.   
At any rate, in the absence of a full
theory of quantum gravity, we are certainly free to postulate such
an effect and to investigate its consequences.

Even in the presence of final-state projection, without further assurances
that go beyond the H-M model the escape of quantum
information from a black hole is by no means certain.
Gottesman and Preskill [5] point out that if the incoming
Hawking radiation interacts with the collapsing matter within the
black hole (as is likely), then the H-M model no longer preserves
quantum information.  In particular, 
let the interaction between incoming Hawking
radiation and matter be given by a unitary transformation $U$.
The transformation between the state of the collapsing
matter and the state of the outgoing Hawking radiation is then
$$ T= { }_{matter\otimes in}\langle \phi| U| \phi\rangle_{in\otimes out}. 
\eqno(4)$$  
Gottesman and Preskill note that if all $U$'s are allowed,
$T$ can be {\it any} matrix satisfying
 $\sum_{m,n} |\langle m| T |n\rangle |^2 = 1$, including transformations
that completely destroy the quantum information in the matter, leading
to purely thermal Hawking radiation.  In general, if the state
$  U |\phi\rangle_{matter\otimes in}$
is not perfectly entangled, then some quantum information in the matter
is lost.

For the purposes of using a black hole as a quantum computer, the key
question is how much quantum information is lost on average
due to such interactions.  I'll now show that for {\it any} final
state, not just the special H-M states, and for {\it almost any} $U$, 
classical information escapes from the hole with certainty,
and quantum information escapes from the hole with fidelity 
$\approx (8/3\pi)^2 \approx .85$.  Essentially, all but half a
qubit of the quantum information escapes.   This fidelity holds in the
limit $N>>1$ and is independent of the exact number of bits 
escaping from the hole: it is the fidelity of escape for the
entire state of the collapsing matter.  Individual quantum bits
inside the hole escape with higher fidelity.  In the limit $N>>1$,
individual quantum bits escape from the hole with fidelity
arbitrarily close to 1.

Let $|\phi\rangle_{matter\otimes in}$ be any final state,
including a product state, and let $U$ be a random unitary
transformation on the matter and incoming Hawking radiation, selected
according to the Haar measure.  (The Haar measure is the unique measure 
over $U(n)$ that is invariant with respect to unitary transformation.) 
In particular, the
final state could be the as yet unknown correct final state specified by the
as yet unknown correct theory of quantum gravity.  Because $U$ is
selected according to the Haar measure, the state
$$|\psi\rangle_{matter\otimes in} = 
U|\phi\rangle_{matter\otimes in} \eqno(5)$$
is a random pure state of the matter and incoming Hawking radiation, i.e.,
a pure state selected according to the uniform measure on the sphere
in $N^2$ dimensions.  That is, it is a random state selected according
to the Hilbert-Schmidt measure.  
The random nature of $U$ implies that the escape
of quantum information from a black hole does not depend on details
of the final state.

Because $|\psi\rangle_{matter\otimes in}$ is random, it is not perfectly
entangled.  As a result, black hole evaporation will not preserve
all the quantum information in the collapsing matter.  But by the same
token, because $|\psi\rangle_{matter\otimes in}$ is random, it is almost
perfectly entangled for large $N$.  In particular, a typical random
state is within one half a qubit of maximum entanglement.

More precisely, a random state in $H_{matter}\otimes H_{in}$ can be written
in Schmidt form as 
$$|\psi\rangle_{matter\otimes in}
= \sum_\ell \lambda_\ell |\ell\rangle'_{matter}
\otimes |\ell\rangle'_{in}.
\eqno(6)$$ 
The distribution of the Schmidt coefficients 
$\lambda_\ell$ for random states is known [17-19].
A random state is almost perfectly entangled [20-21]: the average entropy of
entanglement, $-\sum_\ell \lambda_\ell^2 \log_2
\lambda_\ell^2$, is within one half bit of its maximum possible
value, $\log_2 N$.  It is the high entanglement of random states
that leads to the escape of information from the hole.

We now can calculate the average fidelity with which a state
for the collapsing matter fields
$$|\mu\rangle_{matter} = \sum_\ell 
\mu_\ell |\ell\rangle'_{matter}\eqno(7)$$ 
is transferred to the outgoing Hawking radiation.  

First, look at what happens to the information inside the
hole under final state projection.
Action of $U$ on $|\mu\rangle$
together with the incoming Hawking radiation, followed by
projection onto the final state 
$|\phi\rangle_{matter\otimes in}$, yields a transformation
from the matter to the outgoing Hawking radiation
$$ T= { }_{matter\otimes in}\langle \psi| \phi\rangle_{in\otimes out} 
\eqno(8)$$
The (unnormalized) state of the outgoing Hawking radiation is  
$$|\phi\rangle_{out} = 
{1\over\sqrt N} \sum_\ell \lambda_\ell
\mu_\ell |\ell\rangle'_{out}, \eqno(9)$$
where $\{ |\ell\rangle'_{out} \}$ is a basis
for the Hilbert space of outgoing Hawking radiation, related to the
basis  $\{ |\ell\rangle'_{matter} \}$ 
for the Hilbert space for the collapsing matter via a unitary
transformation $T'$.  Because the normalization of this state depends
in a nonlinear fashion on the $\mu_\ell$, this is a nonlinear
transformation of the input state of the matter.  

Now that we know what happens to quantum states under final state
projection, we can determine what happens to the information inside
the hole. 
Start by looking at the escape of classical information from the hole.
If this information is encoded in the states $|\ell\rangle'_{matter}$,
then the perfect correlation embodied in the Schmidt decomposition,
combined with the nonlinear effect of the final state projection
implies that {\it all} classical information escapes from the hole,
down to the last bit.  Interestingly, the complete escape of
classical information from the hole does not depend on the precise
distribution of the Schmidt coefficients $\lambda_\ell$: it only
requires that they all be non-zero, which occurs with probability
equal to 1.  Properly encoded, classical information escapes from the
hole with certainty.

Now look at how quantum information escapes from the hole.
Comparing the (normalized) outgoing state of the Hawking
radiation with $T'$ times the state of the collapsing matter,
we obtain
$$|{}_{out}\langle \phi| T' |\mu \rangle_{matter} |^2 = 
(\sqrt N \sum_\ell \lambda_\ell |\mu_\ell|^2 )^2. \eqno(10)$$
Since a typical state has $|\mu_\ell|^2 \approx 1/N$,
the state of the collapsing matter is transferred to the
state of the outgoing Hawking radiation with a fidelity
$$f \approx ({1\over\sqrt N} \sum_\ell \lambda_\ell)^2. \eqno(11)$$

This approximate result can be confirmed using standard treatments
of teleportation with imperfectly entangled states [22].
The maximum mean teleportation fidelity attainable using imperfectly
entangled states with Schmidt coefficients $\lambda_\ell$ 
is 
$$\bar f = {1\over N+1} \bigg[ 1 + \big( \sum_\ell \lambda_\ell )^2 \bigg].
\eqno(12)$$
This fidelity is attained for the standard teleportation protocols.
Because escape from a black hole via final state projection is
equivalent to teleportation with a fixed measurement outcome, this is
also the mean fidelity for escape from a black hole.

The techniques of [19] now allow us to estimate the value of
$\bar f$.  for $N >> 1$, we have
$$\langle \sum_\ell \lambda_\ell \rangle 
= \sqrt N { \Gamma(2)\over \Gamma(3/2) \Gamma(5/2)}
\big( 1+ O({1\over N}) \big) \approx {8\over 3\pi} \sqrt N. \eqno(13)$$
As a result, for $N>>1$, we have
$$\bar f \approx \big( {8\over 3\pi} \big)^2 \approx .85 \quad. \eqno(14)$$
Quantum information escapes from the hole with fidelity $\approx .85$.
(Note that in this estimate we are approximating
$\langle \big( \sum_\ell \lambda_\ell )^2 \rangle$ by
$\langle \big( \sum_\ell \lambda_\ell )\rangle^2$ in the limit
that $N>>1$.) 

This fidelity is the fidelity for escape of the entire state of the
collapsing matter.  The fidelity of escape of individual quantum bits
is higher and approaches 1 asymptotically as $N$ becomes large.
Because the escape fidelity lies above the threshold required for quantum
error correction [23], suitably encoded quantum information escapes
from the black hole with fidelity arbitrarily close to 1. 
Given final-state projection, escape from a black hole is almost certain.

Note that in the above demonstration of almost certain escape from
a black hole via final state projection relies on a random
interaction between the collapsing matter and the incoming Hawking radiation.
As only a finite
proper time exists for interaction between the matter and the incoming
Hawking radiation, this interaction is not truly random.
What is important for the escape of the quantum information is not true
randomness, however, but entanglement.  We have recently demonstrated
both theoretically and experimentally that {\it pseudorandom} 
states and transformations, implemented by quantum logic circuits with
$O(2n)$ gates for $n= \log_2 N$ qubits, exhibit the same Schmidt coefficient
statistics as true random states and transformations [24].
Accordingly, we may reasonably hope that the final state projection,
whatever it is, is sufficiently entangled to give high fidelity
transfer of the state of the matter within the hole to the state of
the outgoing Hawking radiation.

The results of this paper suggest that if black holes evaporate
via final state projection, they might make good quantum computers. 
The fidelity of transfer of quantum information is 
better than what is required for robust quantum computation. 
Indeed, if all one wants is a Yes/No answer from the 
computation, i.e., a classical bit,
then the black hole can deliver the answer with certainty.

Note, however, that for 
information to escape from the hole under the final
state projection model, the time evolution apart from the final projection
must remain unitary as the densities of matter and energy approach
the Planck scale near the singularity.  That is, the strategy for escaping
from a black hole presented here assumes that the only source of nonlinearity 
is the final state projection.  Even if the time evolution apart
from the projection is unitary,
a person outside the hole must know the exact interaction that occurred
between the collapsing matter and the incoming Hawking radiation in order
to reconstruct the information escaping from the hole.
Final state projection will
have to await experimental and theoretical confirmation before
black holes can be used as quantum computers.
It would be premature to jump into a black hole
just now.  

\vfill
\noindent{$^*$ slloyd@mit.edu}

\bigskip
\noindent{\it Acknowledgements:}  The author would like to thank
A. Hosoya for bringing this issue to his attention.

\vfil\eject

\bigskip
\noindent{\it References:}

\noindent [1] S. Lloyd,
{\it Nature} {\bf 406}, 1047-1054, 2000

\noindent [2] S. Lloyd, 
       {\it Phys. Rev. Lett.} {\bf 88} (23): art. no. 237901, 2002.

\noindent [3] Y.J. Ng, {\it Phys. Rev. Lett.} {\bf 86}
 (2001) 2946-2949; Erratum-ibid. {\bf 88} (2002) 139902

\noindent [4] G.T. Horowitz and J. Maldacena, ``The Black-Hole Final State,''
hep-th/0310281.

\noindent [5] D. Gottesman and J. Preskill, ``Comment on `The black hole final
state,' " hep-th/0311269.

\noindent [6] U. Yurtsever and G. Hockney, ``Causality, entanglement,
and quantum evolution beyond Cauchy horizons,'' quant-ph/0312160.

\noindent [7] U. Yurtsever and G. Hockney, ``Signaling and the Black Hole Final
State," quant-ph/0402060.

\noindent [8] S. Weinberg, {\it Phys. Rev. Lett.}  {\bf 62}, 485 (1989). 

\noindent [9] J. Polchinksi, {\it Phys. Rev. Lett.} {\bf 66}, pg. 397 (1991).

\noindent [10] N. Gisin, {\it Phys. Lett. A} {\bf 113}, p. 1 (1990).

\noindent [11] A. Peres, {\it Phys. Rev. Lett.} {\bf 63}, 1114 (1989).

\noindent [12] D. Abrams, and S. Lloyd, {\it Phys. Rev. Lett.}, {\bf 81},
3992-3995, (1998).

\noindent [13] P.K. Majumder {\it et. al.},{\it  Phys. Rev. Lett.}
{\bf  65}, 2931 (1990).

\noindent [14] R.L. Walsworth {\it et. al.}, {\it Phys. Rev. Lett.}
{\bf 64}, 2599 (1990).

\noindent [15] T.E. Chupp and R.J. Hoare, {\it Phys. Rev. Lett.} {\bf 64}, 2261
(1990).

\noindent [16] J.J. Bollinger, D.J. Heinzen, W.M. Itano, S.L. Gilbert,
D.J. Wineland, {\it Phys. Rev. Lett.} {\bf 63}, 1031 (1989).

\noindent[17] E. Lubkin, {\it J. Math. Phys.} {\bf 19}, 1028 (1978).

\noindent [18] S. Lloyd and H. Pagels, 
{\it Ann. Phys.} {\bf 188}, 186 (1988). 

\noindent[19] H.-J. Sommers and K. \.Zyczkowski, ``Statistical properties
of random density matrices,'' quant-ph/0405031.

\noindent[20] D.N. Page, {\it Phys. Rev. Lett.} {\bf 71}, 1291 (1993).

\noindent[21] S. Sen, ``Average entropy of a subsystem,'' hep-th/9601132.

\noindent[22] K. Baraszek, Phys. Rev. A62, 024301 (2000); quant-ph/0002088.

\noindent[23] M.A. Nielsen and I.L. Chuang, {\it Quantum Computation
and Quantum Information}, Cambridge University Press, 2003.

\noindent [24] J. Emerson, Y.S. Weinstein, M. Saraceno, S. Lloyd, D. Cory,
{\it Science} {\bf 302}, 2098 (2003). 

\vfill\eject\end